\documentclass[pre,twocolumn,noshowpacs,amsmath,amssymb]{revtex4-2}
\usepackage{graphicx}
\usepackage{amsfonts}
\usepackage{dcolumn}
\usepackage{bm}
\usepackage{color}
\usepackage{hyperref}

\begin{document}
	
\newcommand{\gin}[1]{{\bf\color{blue}#1}}
\def\bc{\begin{center}}
\def\ec{\end{center}}
\def\bea{\begin{eqnarray}}
\def\eea{\end{eqnarray}}
\newcommand{\avg}[1]{\langle{#1}\rangle}
\newcommand{\Avg}[1]{\left\langle{#1}\right\rangle}

\title{Random walks on complex networks under time-dependent stochastic resetting}	
\author{Hanshuang Chen}\email{chenhshf@ahu.edu.cn}
\author{Yanfei Ye}

\affiliation{School of Physics and Optoelectronic Engineering, Anhui University, Hefei 230601, China}

\date{\today}
	
\begin{abstract}	
We study discrete-time random walks on networks subject to a time-dependent stochastic resetting, where the walker either hops randomly between neighboring nodes with a probability $1-\phi(a)$, or is reset to a given node with a complementary probability $\phi(a)$. The resetting probability $\phi(a)$ depends on the time $a$ since the last reset event (also called $a$ the age of the walker). Using the renewal approach and spectral decomposition of transition matrix, we formulize the stationary occupation probability of the walker at each node and the mean first passage time between arbitrary two nodes. Concretely, we consider that two different time-dependent resetting protocols that are both exactly solvable. One is that $\phi(a)$ is a step-shaped function of $a$ and the other one is that $\phi(a)$ is a rational function of $a$. We demonstrate the theoretical results on two different networks, also validated by numerical simulations, and find that the time-modulated resetting protocols can be more advantageous than the constant-probability resetting in accelerating the completion of a target search process.        
\end{abstract}
	\maketitle
	
\section{Introduction}	
Since the seminal work by Evans and Majumdar \cite{evans2011diffusion}, random walks subject to resetting processes have received growing attention in the last decade (see \cite{evans2020stochastic} for a recent review). The walker is stochastically interrupted and reset to the initial position, and the random process is then restarted. Interestingly, the occupation probability at stationarity is strongly altered. The mean  time to reach a given target for the first time can become finite and be minimized with respect to the resetting rate. Some extensions have been made in the field, such as spatially \cite{evans2011diffusion2} or temporally \cite{NJP2016.18.033006,pal2016diffusion,PhysRevE.93.060102,PhysRevE.96.012126,PhysRevE.100.032110} dependent resetting rate,  higher dimensions \cite{Evans2014_Reset_Highd}, complex geometries \cite{Christou2015,PhysRevResearch.2.033027,BressloffJSTAT2021,PhysRevE.105.034109}, noninstantaneous resetting \cite{EvansJPA2018,PalNJP2019,PhysRevE.101.052130,GuptaJPA2020}, in the presence of external potential \cite{pal2015diffusion,ahmad2019first,gupta2020stochastic}, other types of Brownian motion, like run-and-tumble particles \cite{evans2018run,santra2020run,bressloff2020occupation}, active particles \cite{scacchi2018mean,kumar2020active}, constrained Brownian particle \cite{PhysRevLett.128.200603}, and so on \cite{basu2019symmetric}.  
These nontrivial findings have triggered an enormous recent activities in the field, including statistical physics \cite{PhysRevLett.116.170601,pal2017first,gupta2014fluctuating,evans2014diffusion,meylahn2015large,chechkin2018random,PhysRevE.103.022135,de2020optimization,magoni2020ising,arXiv:2202.04906,JPA2022.55.021001,JPA2022.55.234001}, stochastic thermodynamics \cite{fuchs2016stochastic,pal2017integral,gupta2020work}, chemical and biological
processes \cite{reuveni2014role,PhysRevLett.112.240601,rotbart2015michaelis,PhysRevLett.128.148301,JPA2022.55.274005}, extremal statistics \cite{PhysRevE.103.052119,JStatMech2022.063202,JPA2022.55.034002} optimal control theory \cite{arXiv:2112.11416}, and single-particle experiments \cite{tal2020experimental,besga2020optimal}. 

However, the impact of resetting on random walks on networked systems has received only a small amount of attention \cite{avrachenkov2014personalized,avrachenkov2018hitting,christophorov2020peculiarities,PhysRevE.103.012122,PhysRevE.103.052129,huang2021random}. Random walks on complex networks is a simple but very important model \cite{masuda2017random,PhysRevLett.92.118701,PhysRevE.87.012112}. It not only underlies many important dynamical processes on networked systems, such as epidemic spreading \cite{RevModPhys.87.925,colizza2007reaction,PhysRevX.1.011001}, population extinction \cite{WKBReview1,PhysRevLett.117.028302}, neuronal firing \cite{tuckwell1988introduction}, consensus formation \cite{PhysRevLett.94.178701}, 
but also finds a broad range of applications, such as community detection \cite{rosvall2008maps,zhou2004network,pons2005computing}, human mobility \cite{PhysRevE.86.066116,riascos2017emergence,barbosa2018human}, ranking and searching on the web \cite{PhysRevLett.92.118701,newman2005measure,lu2016vital,kleinberg2006complex,RevModPhys.87.1261}. Random walks on networks under resetting have many applications in computer science and physics. For instance, label propagation in machine learning algorithms \cite{Bautista2019}, or the famous PageRank \cite{Pagerank1998}, can be interpreted as a random walker with uniform resetting probability to all the nodes of the network. Human and animal mobility consists of a mixture of short-range moves with intermittent long-range moves where an agent relocates to a new place and then starts local moves \cite{Barabasi.Nature2008,Walsh_NatPhys2010,RevModPhys.83.81}. Until recently, Riascos \textit{et al.} \cite{PhysRevE.101.062147} established relationships between the random walk dynamics with resetting to one node and the spectral representation of the transition matrix in the absence of resetting \cite{rose2018spectral}. Furthermore, they discussed the condition under which resetting becomes advantageous to reduce the mean first passage time (MFPT) \cite{arXiv:2110.15437}. Subsequently, the result was generalized to the case when multiple resetting nodes exist \cite{PhysRevE.103.062126,Chaos2021_31.093135}. In a recent work, we have generalized the constant resetting probability to the case when the resetting probability is node-dependent \cite{JSM2022.053201}.

In this paper we consider a different generalization of the resetting random walks on networks: a time-dependent resetting probability. This generalization is quite natural in the context of target search. When searching for a target, it is unlikely to restart at the beginning. But as time elapses without success, it is more likely to return to the original point and restart the search process. We should note that for continuous-time random walks on one-dimensional space similar problem has been considered in several recent works, including nonexponential waiting times between successive resets \cite{NJP2016.18.033006,PhysRevE.93.060102,PhysRevE.96.012126} and the time-dependent resetting rate \cite{pal2016diffusion}. For the case of a resetting rate that depends on the absolute time elapsed from the beginning of the process was considered in \cite{PhysRevE.100.032110}. In the present work, we focused on discrete-time random walks on arbitrary networks subject to a time-dependent resetting probability $\phi(a)$, where $a$ refers to the time since the last reset event (or we call it the age of the walker), rather than the absolute time from the initial condition. This means that when a reset happens, the walker no longer remembers what happened before resetting, and thus the process is still Markovian. First of all, we formulize the occupation probability distribution and the MFPT for general choice of $\phi(a)$ by the renewal approach combined with the spectrum properties of transition matrix. We then consider two exactly solvable examples for the settings of $\phi(a)$. The first example is that $\phi(a)$ is a step-shaped function where the resetting probability switches from one value to another one at $a=a_c$. The second one is that $\phi(a)=\frac{a}{a+a_c}$ is a strictly increasing function from zero as $a$ increases. The theoretical and simulation results show that such time-modulated resetting protocols are able to expedite the completion of a target search compared with the constant-probability resetting.


\section{Model}\label{sec2}
Consider a walker that performs discrete-time random walks on an undirected and unweighted network of size $N$ \cite{masuda2017random}. At each time step, the walker either hops between two neighboring nodes with a probability $1-\phi(a)$ or is reset to a given node with a complementary probability $\phi(a)$, where $a$ denotes the age of the walker determined by an internal clock carried by the walker itself. For the former, the walker jumps from the current node to one of its neighboring nodes with equal probabilities, in the sense that the transition probability from node $i$ to node $j$ can be written as $W_{ij}=A_{ij}/k_i$, where $A_{ij}$ is the element of adjacency matrix of the underlying network, $k_i=\sum_{j=1}^{N}A_{ij}$ is the degree of node $i$. Meanwhile, the age of the walker is increased by one: $a \to a+1$. For the latter, the walker is reset to the node $r$ (called the resetting node), and its age is simultaneously reinitialized to zero, $a \to 0$. 

\section{Occupation probability}
Let us denote by $P_{ij}(t)$ the probability to find the walker at node $j$ at time $t$ providing it has started from node $i$. $P_{ij}(t)$ satisfies a first renewal equation \cite{pal2016diffusion,ahmad2019first,chechkin2018random,Chaos2021_31.093135}, 
\begin{eqnarray}\label{eq1}
{P_{ij}}\left( t \right) = \Phi \left( t \right){P^{0}_{ij}}\left( t \right) + \sum\limits_{t' = 0}^t {\Psi \left( {t'} \right){P_{rj}}\left( {t - t'} \right)},  
\end{eqnarray}
where 
\begin{eqnarray}\label{eq2}
\Phi \left( t \right) = \prod\limits_{a = 1}^t {\left[ {1 - \phi \left( a \right)} \right]} 
\end{eqnarray}
is the probability of no reset taking place up to time $t$, and 
\begin{eqnarray}\label{eq3}
\Psi \left( t \right) = \left\{ \begin{array}{lr}
\Phi \left( {t - 1} \right)\phi \left( t \right),&t \ge 1\\
0,&t = 0
\end{array} \right.
\end{eqnarray}
is the probability of the first reset taking place at time $t$.
$P^{0}_{ij}(t)$ is the occupation probability of the walker in the absence of resetting processes \cite{PhysRevLett.92.118701}, given by (see Appendix \ref{app1} for details) 
\begin{eqnarray}
{P^{0}_{ij}}\left( t \right) =  \sum\limits_{\ell = 1}^N {\lambda_\ell^t \langle i | {{\phi _\ell}} \rangle } \langle {{{\bar \phi }_\ell}} | j\rangle ,
\end{eqnarray}
where $\lambda_\ell$ is the $\ell$th eigenvalue of the transition matrix $\bm{W}$, and the corresponding left eigenvector and right eigenvector are respectively $\langle {{\bar \phi }_\ell}|$ and $| {\phi_\ell}\rangle$, satisfying $\langle {{{\bar \phi }_\ell}} | {{\phi _m}} \rangle  = {\delta _{\ell m}}$ and $\sum_{\ell=1}^{N} |\phi_{\ell} \rangle \langle {\bar \phi}_{\ell}|=\bm{I}$. $\left| i \right\rangle $ denotes the canonical base with all its components equal to 0 except the $i$th one, which is equal to 1.
The first term on the r.h.s. of Eq.(\ref{eq1}) accounts for the walker is never reset up to time $t$, and the second term accounts for the walker is reset at time $t'$ for the first time, after which the process starts anew from the resetting node $r$ for the remaining time $t-t'$.

Let ${\kappa _{ij}}\left( t \right) = \Phi \left( t \right){P^{0}_{ij}}\left( t \right)$, and take the discrete-time Laplace transform for Eq.(\ref{eq1}), $\tilde f\left( s \right) = \sum\nolimits_{t = 0}^\infty  {{e^{ - st}}f\left( t \right)}$, which yields
\begin{eqnarray}\label{eq4}
{\tilde P_{ij}}\left( s \right) = {\tilde \kappa _{ij}}\left( s \right) + \tilde \Psi \left( s \right){\tilde P_{rj}}\left( s \right).
\end{eqnarray}
Letting $i=r$ in Eq.(\ref{eq4}), we have 
\begin{eqnarray}\label{eq5}
{{\tilde P}_{rj}}\left( s \right) = \frac{{{{\tilde \kappa }_{rj}}\left( s \right)}}{{1 - \tilde \Psi \left( s \right)}}.
\end{eqnarray}
Subsitituting Eq.(\ref{eq5}) into Eq.(\ref{eq4}), we obtain
\begin{eqnarray}\label{eq6}
{{\tilde P}_{ij}}\left( s \right) = {{\tilde \kappa }_{ij}}\left( s \right) + \frac{{\tilde \Psi \left( s \right)}}{{1 - \tilde \Psi \left( s \right)}}{{\tilde \kappa }_{rj}}\left( s \right).
\end{eqnarray}
If the resetting node is the same as the original node, $r=i$, Eq.(\ref{eq6}) simplifies to
\begin{eqnarray}\label{eq7}
{\tilde P_{ij}}\left( s \right) = \frac{{{{\tilde \kappa }_{ij}}\left( s \right)}}{{1 - \tilde \Psi \left( s \right)}}.
\end{eqnarray}
By inverting Eq.(\ref{eq6}), one obtains the occupation probability  $P_{ij}(t)$. However, in most of instances the inverse transform of Eq.(\ref{eq6}) is almost impossible to reach. Instead, one can take the limit, 
\begin{eqnarray}\label{eq7a}
{P_{j}}\left( \infty  \right) = \mathop {\lim }\limits_{s \to 0} \left( {1 - {e^{ - s}}} \right){{\tilde P}_{ij}}\left( s \right) 
\end{eqnarray} 
to obtain the stationary occupation probability of the walker at each node.

\section{Survival probability}
Let us suppose that there is a trap located at node $j$. Once it arrives at the trap, the walker will be absorbed immediately. Let us denote by $Q_{ij}(t)$ the survival probability of the walker at time $t$, providing that the walker has started from node $i$. $Q_{ij}(t)$ satisfies a first renewal equation \cite{pal2016diffusion,ahmad2019first,chechkin2018random,Chaos2021_31.093135},  
\begin{eqnarray}\label{eq8}
{Q_{ij}}\left( t \right) =&& \Phi \left( t \right)Q_{ij}^0\left( t \right)  + (1-\delta_{jr}) \nonumber \\ \times && \sum\limits_{t' = 1}^t {\Psi \left( {t'} \right)Q_{ij}^0\left( {t'-1} \right){Q_{rj}}\left( {t - t'} \right)} ,
\end{eqnarray}
where $Q_{ij}^0\left( t \right)$ denotes the survival probability in the absence of resetting processes (see Appendix \ref{app2} for details). The first term on the r.h.s. of Eq.(\ref{eq8}) corresponds to the case where
there is no resetting event at all up to time $t$, which occurs with probability $\Phi(t)$. The second term accounts for the event where the first resetting takes place at time $t'$, which occurs with probability $\Psi(t')$. Before the first resetting, the walker survives with probability $Q_{ij}^0\left( {t'-1} \right)$, after which the walker survives with probability ${Q_{rj}}\left( {t - t'} \right)$. If the resetting node is the same as the trap node, $r=j$, the walker is immediately absorbed as soon as it is reset. Therefore, the prefactor $1-\delta_{jr}$ ensures the second term on the r.h.s. of Eq.(\ref{eq8}) vanishes when $r=j$.

Let ${\chi _{ij}}\left( t \right) = \Phi \left( t \right)Q_{ij}^0\left( t \right)$, ${\eta _{ij}}\left( t \right) = \Psi \left( t \right)Q_{ij}^0\left( t-1 \right)$, (noting that $\Psi (0)=0$) and take the Laplace transform for Eq.(\ref{eq8}), which yields, 
\begin{eqnarray}\label{eq9}
{{\tilde Q}_{ij}}\left( s \right) = {{\tilde \chi }_{ij}}\left( s \right) +(1-\delta_{jr}) {{\tilde \eta }_{ij}}\left( s \right){{\tilde Q}_{rj}}\left( s \right).
\end{eqnarray}
Letting $i=r$ in Eq.(\ref{eq9}), we have 
\begin{eqnarray}\label{eq10}
{{\tilde Q}_{rj}}\left( s \right) = \frac{{{{\tilde \chi }_{rj}}\left( s \right)}}{{1 - (1-\delta_{jr}) {{\tilde \eta }_{rj}}\left( s \right)}}	.
\end{eqnarray}
Substituting Eq.(\ref{eq10}) into Eq.(\ref{eq9}), we obtain
\begin{eqnarray}\label{eq11}
{{\tilde Q}_{ij}}\left( s \right) = {{\tilde \chi }_{ij}}\left( s \right) + \frac{{\left( {1 - {\delta _{jr}}} \right){{\tilde \eta }_{ij}}\left( s \right)}}{{1 - \left( {1 - {\delta _{jr}}} \right){{\tilde \eta }_{rj}}\left( s \right)}}{{\tilde \chi }_{rj}}\left( s \right).
\end{eqnarray}
If the resetting node coincides with the original node, $r=i$, Eq.(\ref{eq11}) can be simplified as
\begin{eqnarray}\label{eq12}
{{\tilde Q}_{ij}}\left( s \right) = \frac{{{{\tilde \chi }_{ij}}\left( s \right)}}{{1 - (1-\delta_{ij}) {{\tilde \eta }_{ij}}\left( s \right)}}	.
\end{eqnarray}

The MFPT from node $i$ to node $j$ is given by
\begin{eqnarray}\label{eq13}
\left\langle {{T_{ij}}} \right\rangle  &=& {\tilde Q_{ij}}(0)  \nonumber \\ &=& {\tilde \chi _{ij}}(0) + \frac{{\left( {1 - {\delta _{jr}}} \right){{\tilde \eta }_{ij}}(0)}}{{1 - \left( {1 - {\delta _{jr}}} \right){{\tilde \eta }_{rj}}(0)}}{\tilde \chi_{rj}}(0) . 
\end{eqnarray}
For the case when the resetting node is the same as the original node, $r=i$, Eq.(\ref{eq13}) simplifies to 
\begin{eqnarray}\label{eq14}
\left\langle {{T_{ij}}} \right\rangle  =  \frac{{\tilde \chi_{ij}}(0)}{{1 - \left( {1 - {\delta _{ij}}} \right){{\tilde \eta }_{ij}}(0)}} .
\end{eqnarray}

\section{Constant-probability resetting}
For completeness we first consider the case when the resetting probability at each time step is a constant, $\phi(a)=\gamma$, that is independent of the age of the walker. In this case, we have $\Phi(t) = (1 - \gamma)^{t}$, $\Psi\left(t\right)=\left( {1 - \gamma } \right)^{t-1}\gamma$ for $t \ge 1$, $\Psi(0)=0$ for $t=0$,
\begin{eqnarray}\label{eq2.1}
{\kappa _{ij}}\left( t \right) = \Phi \left( t \right){P^{0}_{ij}}\left( t \right) = \sum\limits_{\ell = 1}^N {\lambda _\ell^t {{\left( {1 - \gamma } \right)}^{t }} \langle i | {\phi_\ell}\rangle } \langle {{\bar \phi }_\ell} | j \rangle .
\end{eqnarray}
Taking the Laplace transform for $\Psi(t)$  and $\kappa_{ij}(t)$, we have
\begin{eqnarray}\label{eq2.2}
\tilde \Psi \left( s \right) = \frac{\gamma e^{ - s} }{{1 - \left( {1 - \gamma } \right){e^{ - s}}}}
\end{eqnarray}
and 
\begin{eqnarray}\label{eq2.3}
{\tilde \kappa _{ij}}\left( s \right) = \sum\limits_{\ell = 1}^N {\frac{{1  }}{{1 - \lambda _\ell^{}\left( {1 - \gamma } \right){e^{ - s}}}}\langle i | {\phi_\ell}\rangle } \langle {{\bar \phi }_\ell} | j \rangle.
\end{eqnarray}
Substituting Eq.(\ref{eq2.2}) and Eq.(\ref{eq2.3}) into Eq.(\ref{eq6}), we have
\begin{eqnarray}\label{eq2.4}
{\tilde P_{ij}}\left( s \right) &=& \frac{{\langle { {\bar{\phi}_1} |j}\rangle }}{{1 - {e^{ - s}}}} + \sum\limits_{\ell = 2}^N {\frac{1}{{1 - {\lambda _\ell}\left( {1 - \gamma } \right){e^{ - s}}}}} \langle { i |{\phi_\ell}} \rangle \langle { {\bar{\phi}_\ell} |j} \rangle  \nonumber \\ &+& \frac{{\gamma {e^{ - s}}}}{{1 - {e^{ - s}}}}\sum\limits_{\ell = 2}^N {\frac{1}{{1 - {\lambda _\ell}\left( {1 - \gamma } \right){e^{ - s}}}}} \langle { r |{\phi_\ell}} \rangle \langle { {\bar {\phi} }_\ell |j} \rangle 
\end{eqnarray}
Taking the inverse transform for Eq.(\ref{eq2.4}), we have 
\begin{eqnarray}\label{eq2.5}
P_{ij}(t) &=& \langle { {\bar {\phi}_1}|j} \rangle  + \sum\limits_{\ell = 2}^N {\lambda _\ell^t{{\left( {1 - \gamma } \right)}^t}} \langle { i |{\phi_\ell}}\rangle \langle { {{{\bar \phi }_\ell}} |j} \rangle \nonumber \\ & +& \gamma \sum\limits_{\ell = 2}^N {\frac{{1-\lambda_\ell^t{{\left( {1 - \gamma } \right)}^t}}}{{1 - {\lambda _\ell}\left( {1 - \gamma } \right)}}} \langle { r |{\phi _\ell}} \rangle \langle { {{{\bar \phi }_\ell}} |j} \rangle 
\end{eqnarray}
In stationarity, $t \to \infty$, $\lambda_\ell^t \to 0$  for $\ell=2,\cdots,N$, we get to the stationary occupation probability in the presence of constant-probability resetting processes,
\begin{eqnarray}\label{eq2.6}
{P_{j}}\left( \infty  \right) = \langle {{{\bar \phi }_1}} | j \rangle  + \gamma \sum\limits_{\ell = 2}^N {\frac{{\langle r | {{\phi _\ell}} \rangle \langle {{{\bar \phi }_\ell}} | j \rangle }}{{1 - \lambda _\ell^{}\left( {1 - \gamma } \right)}}} .
\end{eqnarray}
The first term on r.h.s. of Eq.(\ref{eq2.6}) is stationary occupation probability in the absence of resetting \cite{PhysRevLett.92.118701}, and the second term is an nonequilibrium contribution due to the resetting processes.

In the following, we will derive the MFPT for the case of a constant resetting probability. To the end, we take the Laplace transform for $\chi_{ij}(t)$ and $\eta_{ij}(t)$, which yields
\begin{eqnarray}\label{eq2.7}
{{\tilde \chi }_{ij}}\left( s \right)& = &\sum\limits_{t = 0}^\infty  {{e^{ - st}}\Phi \left( t \right)Q_{ij}^0\left( t \right)}  = \sum\limits_{t = 0}^\infty  {{e^{ - s't}}Q_{ij}^0\left( t \right)}  \nonumber \\  &=& \tilde Q_{ij}^0\left( {s'} \right),
\end{eqnarray}
and 
\begin{eqnarray}\label{eq2.8}
{{\tilde \eta }_{ij}}\left( s \right) &=& \sum\limits_{t = 1}^\infty  {{e^{ - st}}\Psi \left( t \right)Q_{ij}^0\left( t-1 \right)}  = \gamma e^{-s} \sum\limits_{t = 0}^\infty  {{e^{ - s't}}Q_{ij}^0\left( t \right)}  \nonumber \\  &=& \gamma e^{-s} \tilde Q_{ij}^0\left( {s'} \right),
\end{eqnarray}
where $s' = s - \ln \left( {1 - \gamma } \right)$. Substituting Eq.(\ref{eq2.7}) and Eq.(\ref{eq2.8}) into Eq.(\ref{eq11}), we have 
\begin{eqnarray}\label{eq2.9}
{{\tilde Q}_{ij}}\left( s \right) = \tilde Q_{ij}^0\left( {s'} \right) + \frac{{\left( {1 - {\delta _{jr}}} \right)\gamma {e^{ - s}}\tilde Q_{ij}^0\left( {s'} \right)}}{{1 - \left( {1 - {\delta _{jr}}} \right)\gamma {e^{ - s}}\tilde Q_{rj}^0\left( {s'} \right)}}\tilde Q_{rj}^0\left( {s'} \right). \nonumber \\ 
\end{eqnarray}

The MFPT is given by Eq.(\ref{eq13}) combined with Eq.(\ref{eq2.9}),
\begin{eqnarray}\label{eq2.10}
\left\langle {{T_{ij}}} \right\rangle & =& \tilde Q_{ij}^0\left( { - \ln \left( {1 - \gamma } \right)} \right)  \nonumber \\  &+& \frac{{\left( {1 - {\delta _{jr}}} \right)\gamma \tilde Q_{ij}^0\left( { - \ln \left( {1 - \gamma } \right)} \right)}}{{1 - \left( {1 - {\delta _{jr}}} \right)\gamma \tilde Q_{rj}^0\left( { - \ln \left( {1 - \gamma } \right)} \right)}} \nonumber \tilde Q_{rj}^0\left( { - \ln \left( {1 - \gamma } \right)} \right).\nonumber \\ 
\end{eqnarray}
In terms of Eq.(\ref{a2.5}), we have 
\begin{eqnarray}\label{eq2.11}
\tilde Q_{ij}^0\left( { - \ln \left( {1 - \gamma } \right)} \right) = \frac{{\sum\limits_{\ell = 2}^N {\frac{{\langle j | {{\phi _\ell}} \rangle \langle {{{\bar \phi }_\ell}} | j \rangle  - \langle i | {{\phi _\ell}} \rangle \langle {{{\bar \phi }_\ell}} | j \rangle }}{{1 - {\lambda _\ell}\left( {1 - \gamma } \right)}}}  + {\delta _{ij}}}}{{\langle {{{\bar \phi }_1}} | j \rangle  + \gamma \sum\limits_{\ell = 2}^N {\frac{{\langle j | {{\phi _\ell}} \rangle \langle {{{\bar \phi }_\ell}} | j \rangle }}{{1 - {\lambda _\ell}\left( {1 - \gamma } \right)}}} }}.\nonumber \\ 
\end{eqnarray}
Substituting Eq.(\ref{eq2.11}) into Eq.(\ref{eq2.10}) and combining Eq.(\ref{eq2.6}), we have
\begin{eqnarray}\label{eq2.12}
\left\langle {{T_{ij}}} \right\rangle =
\frac{1}{{{P_{j}}\left( \infty  \right)}}\left[ {\sum\limits_{\ell = 2}^N {\frac{{\langle j | {{\phi _\ell}} \rangle \langle {{{\bar \phi }_\ell}} | j \rangle  - \langle i | {{\phi _\ell}} \rangle \langle {{{\bar \phi }_\ell}} | j \rangle }}{{1 - {\lambda _\ell}\left( {1 - \gamma } \right)}}}  + {\delta _{ij}}} \right]. \nonumber \\ 
\end{eqnarray}
This is consistent with the results of \cite{PhysRevE.101.062147}.

\section{Time-dependent resetting} 
\subsection{Step-shaped resetting probability}
We consider the resetting probability as a step-shaped function of $a$, 
\begin{eqnarray}\label{eq3.1}
\phi \left( a \right) = \left\{ \begin{array}{lr}
{\gamma _1}, &a \le {a_c},\\
{\gamma _2},& a > {a_c},
\end{array} \right.
\end{eqnarray}
where $a_c$ is a characteristic age that controls the time when the value of the resetting probability is switched. According to Eq.(\ref{eq2}) and Eq.(\ref{eq3}), we have 
\begin{eqnarray}\label{eq3.2}
\Phi \left( t \right) = \left\{ \begin{array}{lr}
{\left( {1 - {\gamma _1}} \right)^t}, & t \le {a_c},\\
{\left( {1 - {\gamma _1}} \right)^{{a_c}}}{\left( {1 - {\gamma _2}} \right)^{t - {a_c}}},&t > {a_c},
\end{array} \right.
\end{eqnarray}
and 
\begin{eqnarray}\label{eq3.3}
\Psi \left( t \right) = \left\{ \begin{array}{lr}
{\left( {1 - {\gamma _1}} \right)^{t - 1}}{\gamma _1}, &t \le {a_c},\\
{\left( {1 - {\gamma _1}} \right)^{{a_c}}}{\left( {1 - {\gamma _2}} \right)^{t - {a_c} - 1}}{\gamma _2},&t > {a_c}, 
\end{array} \right.
\end{eqnarray}
for $t \geq 1$, and $\Psi(0)=0$. 
Performing the Laplace transform for $\Psi(t)$ and $\kappa_{ij}(t)=\Phi(t) P^0_{ij}(t) $, we have 
\begin{eqnarray}\label{eq3.4}
\tilde \Psi \left( s \right) &= &\frac{{{\gamma _1}{e^{ - s}}\left[ {1 - {e^{ - s{a_c}}}{{\left( {1 - {\gamma _1}} \right)}^{{a_c}}}} \right]}}{{1 - \left( {1 - {\gamma _1}} \right){e^{ - s}}}} \nonumber \\ & +& \frac{{{\gamma _2} {{e^{ - s{(a_c+1)}}}{{\left( {1 - {\gamma _1}} \right)}^{{a_c}}}} }}{{1 - \left( {1 - {\gamma _2}} \right){e^{ - s}}}}
\end{eqnarray}
and 
\begin{eqnarray}\label{eq3.5}
{{\tilde \kappa }_{ij}}(s) = \sum\limits_{\ell = 1}^N \left[ \frac{{1 - {e^{ - s(1+{a_c})}}\lambda _\ell^{1+{a_c} }{{\left( {1 - {\gamma _1}} \right)}^{1+{a_c} }}}}{{1 - {\lambda _\ell}\left( {1 - {\gamma _1}} \right){e^{ - s}}}} \right. \nonumber \\ \left. +  \frac{{{e^{ - s(1+{a_c})}}\lambda _\ell^{1+{a_c} }{{\left( {1 - {\gamma _1}} \right)}^{{a_c}}}\left( {1 - {\gamma _2}} \right)}}{{1 - {\lambda _\ell}\left( {1 - {\gamma _2}} \right){e^{ - s}}}} \right] \langle { i |{\phi _\ell}} \rangle \langle { {{{\bar \phi }_\ell}} |j} \rangle 
\end{eqnarray}
Substituting Eq.(\ref{eq3.4}) and Eq.(\ref{eq3.5}) into Eq.(\ref{eq7a}), we obtain the stationary occupation probability, 
\begin{widetext}
\begin{eqnarray}\label{eq3.6}
{P_{j}}\left( \infty  \right)= \langle { {{{\bar \phi }_1}} |j} \rangle  + \sum\limits_{\ell = 2}^N {\frac{{{\gamma _1}{\gamma _2}\left[ {1 - \left( {1 - {\gamma _2}} \right){\lambda _\ell} + {{\left( {1 - {\gamma _1}} \right)}^{{a_c}}}\left( {{\gamma _1} - {\gamma _2}} \right)\lambda _\ell^{{a_c} + 1}} \right]}}{{\left[ {1 - \left( {1 - {\gamma _1}} \right){\lambda _\ell}} \right]\left[ {1 - \left( {1 - {\gamma _2}} \right){\lambda _\ell}} \right]\left[ {{{\left( {1 - {\gamma _1}} \right)}^{{a_c}}}\left( {{\gamma _1} - {\gamma _2}} \right) + {\gamma _2}} \right]}}} \langle { r |{\phi _\ell}} \rangle \langle { {{{\bar \phi }_\ell}} |j} \rangle 
\end{eqnarray}
\end{widetext}
If $\gamma_1=\gamma_2$, Eq.(\ref{eq3.6}) recovers to the result of Eq.(\ref{eq2.6}) when the resetting probability is a constant.

\begin{figure*}
	\centerline{\includegraphics*[width=1.8\columnwidth]{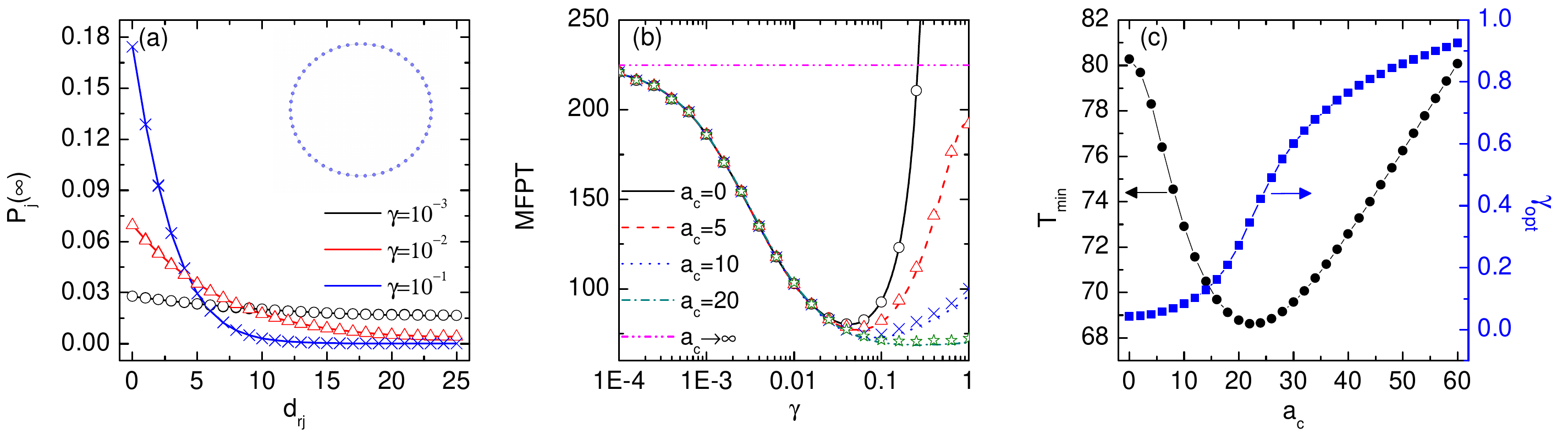}}
	\caption{Results for the step-shaped resetting (see Eq.(\ref{eq3.1}) for $\gamma_1=0$ and $\gamma_2=\gamma$) on a ring network of $N=50$ (see the inset of (a)). The resetting node is set to be the same as the starting node, $r=i$. (a) The stationary occupation probability $P_j(\infty)$ at node $j$ as a function of the distance $d_{rj}$ to the resetting node $r$ for different $\gamma$. The characteristic age is fixed at $a_c=10$. (b) The MFPT as a function of the resetting probability $\gamma$ for different $a_c$. Lines and symbols correspond to the theory and simulation results, respectively. Note that $a_c=0$ corresponds to the case of the resetting with a constant probability $\gamma$, and $a_c\to \infty$ to the case without resetting. (c) The minimum $T_{\min}$ of the MFPT and the corresponding optimal resetting probability $\gamma_{\rm{opt}}$ as a function of $a_c$. The distance between the starting node and the target node is $d_{ij}=5$.  \label{fig1}}
\end{figure*}

In order to obtain the MFPT by Eq.(\ref{eq13}), we need to derive the expression of ${{\tilde \chi }_{ij}}\left( s \right)$ and ${{\tilde \eta }_{ij}}\left( s \right)$ at $s=0$. According to Eq.(\ref{a3.3}), we have
\begin{eqnarray}\label{eq3.9}
{\tilde \chi}_{ij}(0) =\sum\limits_{\ell = 1}^{N } \left[ \frac{{1 - {{\left( {1 - {\gamma _1}} \right)}^{{a_c} + 1}}{{\left[ {\zeta _\ell^{(j)}} \right]}^{{a_c} + 1}}}}{{1 - \left( {1 - {\gamma _1}} \right)\zeta _\ell^{(j)}}} \right. \nonumber \\ \left.  + \frac{{{{\left( {1 - {\gamma _1}} \right)}^{{a_c}}}\left( {1 - {\gamma _2}} \right){{\left[ {\zeta _\ell^{(j)}} \right]}^{{a_c} + 1}}}}{{1 - \left( {1 - {\gamma _2}} \right)\zeta _\ell^{(j)}}} \right] \langle i|\psi_\ell^{(j)} \rangle \langle \bar{\psi}_\ell^{(j)}|\bm{1}\rangle 
\end{eqnarray}
and 
\begin{eqnarray}\label{eq3.10}
{\tilde \eta}_{ij}(0) =\sum\limits_{\ell = 1}^{N } \left[ \frac{{{\gamma _1}\left[ {1-{{\left( {1 - {\gamma _1}} \right)}^{{a_c}}}{{\left[ {\zeta_\ell^{(j)}} \right]}^{{a_c}}}} \right]}}{{1 - \left( {1 - {\gamma _1}} \right)\zeta_\ell^{(j)}}} \right. \nonumber \\ \left. + \frac{{{{\left( {1 - {\gamma _1}} \right)}^{{a_c}}}{\gamma _2}{{\left[ {\zeta_\ell^{(j)}} \right]}^{{a_c}}}}}{{1 - \left( {1 - {\gamma _2}} \right)\zeta_\ell^{(j)}}} \right] \langle i|\psi_\ell^{(j)} \rangle \langle \bar{\psi}_\ell^{(j)}|\bm{1}\rangle 
\end{eqnarray}
for $i \neq j$. Here $| {\bf{1}} \rangle =(1,\cdots,1)^\top$ is an $N$-dimensional right vector with all components equal to one. $\zeta_\ell ^{(j)}$ is the $\ell$th eigenvalue of the matrix $\bm{W_j}$, and the associated left and right eigenvectors are respectively $\langle {\bar \psi _\ell^{(j)}} |$ and $| {\psi _\ell^{(j)}} \rangle$, satisfying $\langle {\bar{\psi}_\ell^{(j)}}  | {\psi _m^{(j)}} \rangle  = {\delta _{\ell m}}$ and $\sum_{\ell=1}^{N} | {\psi _\ell^{(j)}} \rangle \langle  {\bar \psi _\ell^{(j)}}|=\bm{I}$  The matrix $\bm{W_j}$ is obtained by letting all the entries in the $j$th row and the $j$th column of $\bm{W}$ equal to zero (see Appendix \ref{app3} for details).

Considering the following special case: $\gamma_1=0$, $\gamma_2=\gamma$. Eq.(\ref{eq3.6}) is simplified to
\begin{eqnarray}\label{eq3.7}
	{P_j}(\infty )= \langle { {{{\bar \phi }_1}} |j} \rangle + && \gamma \sum\limits_{\ell = 2}^N {\frac{{1 - {\lambda _\ell} + \gamma {\lambda _\ell}\left( {1 - \lambda _\ell^{{a_c}}} \right)}}{{\left( {1 + \gamma {a_c}} \right)\left( {1 - {\lambda _\ell}} \right)\left[ {1 - \left( {1 - \gamma } \right){\lambda _\ell}} \right]}}} \nonumber \\ && \times \langle { r |{\phi _\ell}} \rangle \langle { {{{\bar \phi }_\ell}} |j} \rangle .
\end{eqnarray}
Again, the first term on r.h.s of Eq.(\ref{eq3.7}) corresponds to the stationary distribution of the standard random walk, and the second term to a nonequilibrium contribution due to resetting. For $a_c \to \infty$, the second term in Eq.(\ref{eq3.7}) vanishes and thus recovers to the result when the resetting is absent.   
	
Eq.(\ref{eq3.9}) and Eq.(\ref{eq3.10}) for $\gamma_1=0$ and $\gamma_2=\gamma$ can be simplified to 
\begin{eqnarray}\label{eq3.13}
	{{\tilde \chi }_{ij}}(0)  = && \sum\limits_{\ell = 1}^{N } \left\{ {\frac{{1 - {{\left[ {\zeta _\ell^{\left( j \right)}} \right]}^{1 + {a_c}}}}}{{1 - \zeta _\ell^{\left( j \right)}}} + \frac{{\left( {1 - \gamma } \right){{\left[ {\zeta_\ell^{\left( j \right)}} \right]}^{1 + {a_c}}}}}{{1 - \left( {1 - \gamma } \right)\zeta_\ell^{\left( j \right)}}}} \right\}  \nonumber \\ && \times \langle i|\psi_\ell^{(j)} \rangle \langle \bar{\psi}_\ell^{(j)}|\bm{1}\rangle
\end{eqnarray}
and	
\begin{eqnarray}\label{eq3.14}
	{{\tilde \eta }_{ij}}(0)=  \gamma \sum\limits_{\ell = 1}^{N }  \frac{{{{\left[ {\zeta_\ell^{\left( j \right)}} \right]}^{ {a_c}}}}}{{1 - \left( {1 - \gamma } \right)\zeta_\ell^{\left( j \right)}}} \langle i|\psi_\ell^{(j)} \rangle \langle \bar{\psi}_\ell^{(j)}|\bm{1}\rangle
\end{eqnarray}
We should note that for $a_c=0$ the model recovers to the case of the resetting with constant probability. For $a_c\to \infty$ it corresponds to the standard random walks without resetting. 

\begin{figure*}
	\centerline{\includegraphics*[width=1.8\columnwidth]{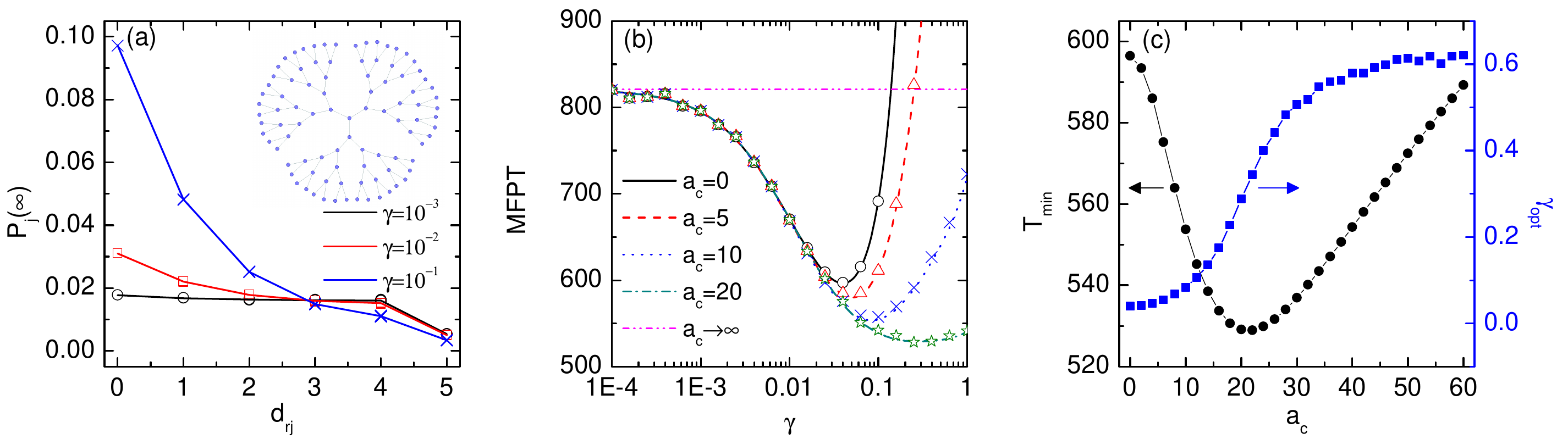}}
	\caption{Results for the step-shaped resetting (see Eq.(\ref{eq3.1}) for $\gamma_1=0$ and $\gamma_2=\gamma$) on a Cayley tree $C_{3,5}$ (see the inset of (a)). The central node is set to be the resetting node. (a) The stationary occupation probability $P_j(\infty)$ at node $j$ as a function of the distance $d_{rj}$ to the resetting node $r$ for different $\gamma$. The characteristic age is fixed at $a_c=10$. (b) The MFPT as a function of the resetting probability $\gamma$ for different $a_c$. Lines and symbols correspond to the theory and simulation results, respectively. Note that $a_c=0$ corresponds to the case of the resetting with a constant probability $\gamma$, and $a_c\to \infty$ to the case without resetting. (c) The minimum $T_{\min}$ of the MFPT and the corresponding optimal resetting probability $\gamma_{\rm{opt}}$ as a function of $a_c$. The starting node is the same as the resetting node, and one of outermost nodes is set to be the target node.  \label{fig2}}
\end{figure*}

To demonstrate the theoretical results, we first consider a ring network with the size $N=50$ (see the inset of In Fig.\ref{fig1}(a)). In Fig.\ref{fig1}(a), we show the stationary occupation probability at each node as a function of the distance to the resetting node for three different resetting probabilities $\gamma$, where the characteristic age is fixed at $a_c=10$. To validate the theoretical results, we also performed the Monte Carlo simulations, and found that there is a good agreement between theory and simulation. The stationary occupation probability decreases monotonically with the distance to the resetting node increases. For a larger resetting probability, there is a larger probability of staying near the resetting position. In Fig.\ref{fig1}(b), we show the MFPT as a function of $\gamma$ for several different values of $a_c$, where the distance between the starting node and the target node is fixed at $d_{ij}=5$, and the resetting node is set to be the same as the starting node. We can see that the MFPT under resetting shows a nonmonotonic dependence on $\gamma$. There exists an optimal resetting probability $\gamma_{\rm{opt}}$ for which the MFPT attains a minimum $T_{\min}$. In a wide range of $\gamma$, the MFPT is less than that for the case without resetting ($a_c \to \infty$), implying that the completion of the first passage process can be expedited by the resetting. On the other hand, such a time-dependent resetting protocol is more advantageous than the constant-probability resetting ($a_c=0$). This is because that the minimum of MFPT is able to become smaller than that for the constant-probability resetting $T_{\min}^{consreset}=80.274$, as shown in Fig.\ref{fig1}(c). The $T_{\min}$ shows a minimum $T_{\min}=68.624$ at $a_c=22$ and $\gamma_{\rm{opt}}=0.346$. The decrease in $T_{\min}$ due to the step-shaped resetting is considerable.

As the second example, we consider a Cayley tree $C_{b,n}$, where $b$ is the coordination number except for the outermost nodes and $n$ is the number of shells. The network is generated as follows. Initially ($n = 0$), $C_{b,0}$ consists of only a central node. To form $C_{b, 1}$, $b$ nodes are created and are attached to the central node. For any $n > 1$, $C_{b, n}$ is obtained from $C_{b, n-1}$ by performing the following operation. For each boundary node of $C_{b, n-1}$, $b-1$ nodes are generated and attached to the boundary node. The size of Cayley tree is $N=1+b(2^n-1)$. The central node is set to be the starting node and one of outermost nodes is set to be the target node. The resetting node is the same as the starting one. In Fig.\ref{fig2}, we show the results on a Cayley tree $C_{3,5}$, and they are similar to those on a ring network.

\subsection{Anti-aging resetting probability}
We consider an anti-aging resetting protocol, where the resetting probability is a strictly increasing function of the age $a$ of the walker. A particular choice is
\begin{eqnarray}\label{eq6.0}
\phi(a)=\frac{a}{a+a_c} ,
\end{eqnarray}
where $a_c>0$ is a parameter that determines the growth rate of the resetting probability with $a$. According to the definitions in Eq.(\ref{eq2}) and Eq.(\ref{eq3}), we get
\begin{eqnarray}\label{eq6.1}
\Phi ( t ) = \prod\limits_{a = 1}^t {\left[ {1 - \frac{a }{a+a_c}} \right]}  =  \frac{ a_c^t {\Gamma \left( { 1 + a_c } \right)}}{{\Gamma \left( {1 +  a_c + t } \right)}}
\end{eqnarray}
and 
\begin{eqnarray}\label{eq6.2}
\Psi ( t ) = \frac{t }{t+a_c}\prod\limits_{a = 1}^{t - 1} {\left[ {1 - \frac{a }{a+a_c}} \right]}  = \frac{t a_c^t \Gamma (a_c) }{\Gamma (1+a_c+t)} 
\end{eqnarray}
where $\Gamma \left( x \right) = \int_0^\infty  {{u^{x - 1}}{e^{ - u}}{\rm{d}}u}$ is the gamma function. Performing Laplace transform for $\Psi(t)$ and $\kappa_{ij}(t) = \Phi(t){P^{0}_{ij}}(t)$, combining the spectral decomposition in Eq.(\ref{a1.2}), we have
\begin{eqnarray}\label{eq6.3}
\tilde \Psi \left( s \right) = {e^{ - s}} + && \left( {1 - {e^{ - s}}} \right){\left( {{a_c}{e^{ - s}}} \right)^{ - {a_c}}}{e^{{a_c}{e^{ - s}} - s}} \nonumber \\ && \times \tilde \Gamma \left( {1 + {a_c},{a_c}{e^{ - s}}} \right) 
\end{eqnarray}
and 
\begin{eqnarray}\label{eq6.4}
{{\tilde \kappa }_{ij}}( s ) =&&  \sum\limits_{\ell = 1}^N {\langle {i| {{\phi _\ell}} } \rangle \langle {{{\bar \phi }_\ell}| j } \rangle a_c {e^{a_c {e^{ - s}}{\lambda _\ell}}}}  \nonumber \\&& \times {\left( {a_c {e^{ - s}}{\lambda _\ell}} \right)^{ - a_c }}  \tilde \Gamma \left( {a_c ,a_c {e^{ - s}}{\lambda _\ell}} \right), 
\end{eqnarray}
where $\tilde \Gamma \left( {x,y} \right) = \int_0^y {{u^{x - 1}}{e^{ - u}}{\rm{d}}u} $  is the lower incomplete gamma function. 

Substituting Eq.(\ref{eq6.3}) and Eq.(\ref{eq6.4}) into Eq.(\ref{eq6}), and then calculating the limit in Eq.(\ref{eq7a}), we obtain the stationary occupation probability of the walker in each node, 
\begin{eqnarray}\label{eq6.5}
{P_j}\left( \infty  \right) = \langle { {{{\bar \phi }_1}} |j} \rangle+ &&   \sum\limits_{\ell = 2}^N \frac{{{e^{{a_c}\left( {{\lambda_\ell} - 1} \right)}} \lambda_\ell^{-a_c} \tilde \Gamma \left( {{a_c},{a_c}{\lambda _\ell}} \right)}}{{\tilde \Gamma \left( {{a_c},{a_c}} \right)}} \nonumber \\ && \times \langle { r |{\phi _\ell}} \rangle \langle { {{{\bar \phi }_\ell}} |j} \rangle. 
\end{eqnarray}
We emphasize again that the second term in Eq(\ref{eq6.5}) is caused by the resetting.

\begin{figure}
	\centerline{\includegraphics*[width=1.0\columnwidth]{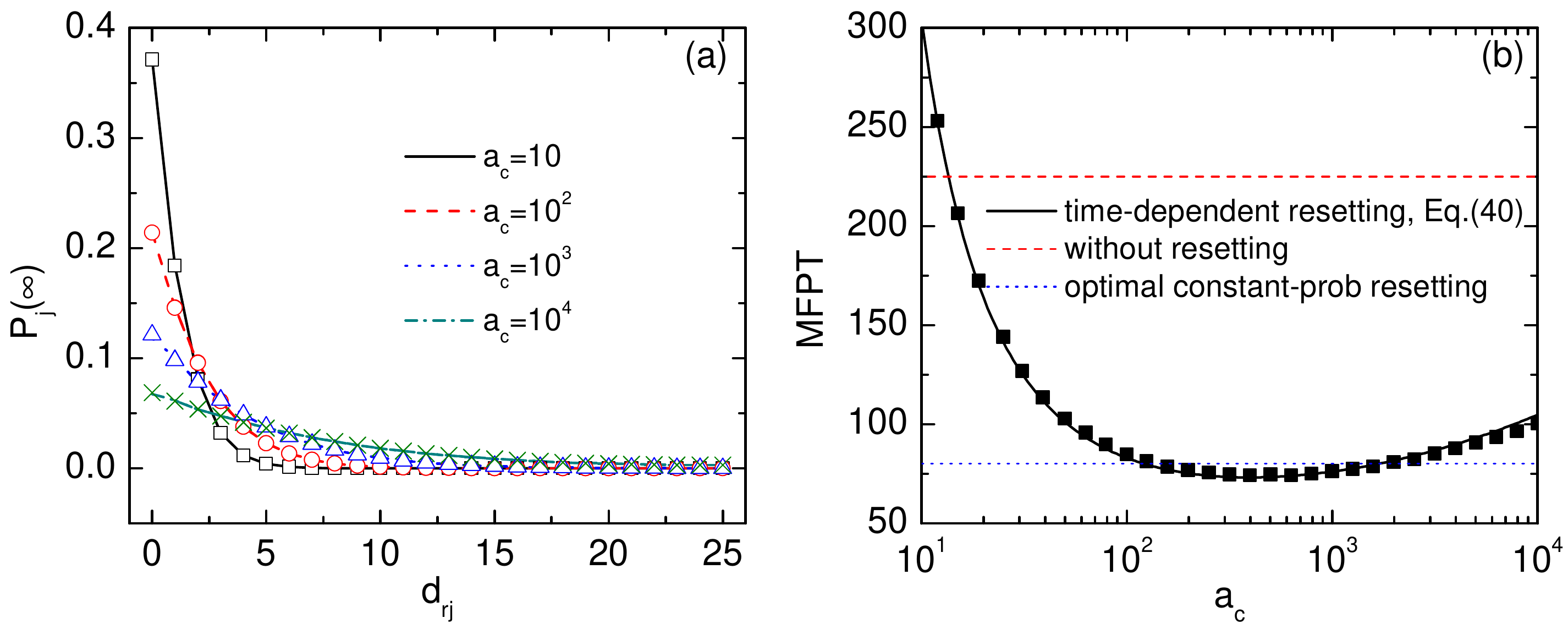}}
	\caption{Results for an anti-aging resetting (see Eq.(\ref{eq6.0})) on a ring network of $N=50$. (a) The stationary occupation probability $P_j(\infty)$ at node $j$ as a function of the distance $d_{rj}$ to the resetting node $r$ for four different value of $a_c$. (b) The MFPT as a function of $a_c$. The resetting node is set to be the same as the starting node, $r=i$. The distance between the starting node and the target node is $d_{ij}=5$.  Lines and symbols correspond to the theory and simulations, respectively. The dashed  horizontal line in (b) indicate the value of MFPT for the case of without resetting, and the dotted horizontal line to the minimum value of MFPT for the case of constant-probability resetting. \label{fig4}}
\end{figure}

To derive the MFPT for node $i$ to node $j$, we compute $\tilde \chi_{ij}( 0 )=\sum\nolimits_{t = 0}^\infty  {\Phi ( t )Q_{ij}^0(t)}$ and $\tilde \eta_{ij}( 0 )=\sum\nolimits_{t = 1}^\infty  {\Psi ( t )Q_{ij}^0(t-1)}$ in terms of Eq.(\ref{eq6.1}), Eq.(\ref{eq6.2}) and Eq.(\ref{a3.3}),
\begin{eqnarray}\label{eq6.6}
{{\tilde \chi }_{ij}}( 0 )=&& \sum\limits_{\ell = 1}^N {{a_c}} {\left( {{a_c}\zeta _\ell^{( j )}} \right)^{ - {a_c}}}{e^{{a_c}\zeta _\ell^{( j )}}}\tilde \Gamma \left( {{a_c},{a_c}\zeta _\ell^{( j )}} \right) \nonumber \\ && \times \langle i|\psi_\ell^{(j)} \rangle \langle \bar{\psi}_\ell^{(j)}|\bm{1}\rangle,
\end{eqnarray}
and
\begin{eqnarray}\label{eq6.7}
{\tilde \eta}_{ij}(0) = \sum\limits_{\ell = 1}^N \left[ 1 +  {{\left( {{a_c}\zeta _\ell^{( j )}} \right)}^{ - {a_c}}}{e^{{a_c}\zeta _\ell^{( j )}}}\left( {1 - 1/\zeta _\ell^{( j )}} \right) \right. \nonumber \\ \left.  \times \tilde \Gamma \left({1 + {a_c},{a_c}\zeta _l^{(j)}} \right) \right] \langle i|\psi_\ell^{(j)} \rangle \langle \bar{\psi}_\ell^{(j)}|\bm{1}\rangle.
\end{eqnarray}
Inserting Eq.(\ref{eq6.6}) and Eq.(\ref{eq6.7}) into Eq.(\ref{eq13}), we obtain the MFPT between arbitrary two nodes. 

In Fig.\ref{fig4}, we compare the theory and simulation results for the anti-aging resetting protocol in Eq.(\ref{eq6.0}) on a ring network of size $N=50$, where the resetting node is set to be the same as the starting node. The theoretical results are in excellent agreement with simulation data. In Fig.\ref{fig4}(a) the stationary occupation probability at each node is plotted for different $a_c$. As expected, the walker has a larger probability to stay near the resetting node for a smaller $a_c$. In Fig.\ref{fig4}(b) we show the MFPT between two nodes with a distance of $d_{ij}=5$ as a function of $a_c$. The MFPT exhibits a nonmonotonic dependence on $a_c$. An optimal $a_c$ appears at $a_c=398$ for which the MFPT admits a minimum $T_{\min} =73.167$. The minimum value of the MFPT is far less than the MFPT without resetting $\langle T_{ij}^{noreset} \rangle =d_{ij}(N-d_{ij})=225$ \cite{huang2021random}, and even slightly less than the minimum of the MFPT for the constant-probability resetting, that is $T_{\min}^{consreset}=80.274$. 

\section{Conclusions}
In conclusion, we have explored discrete-time random walks on networks subject to a time-dependent resetting probability to a given node. Here the resetting probability $\phi(a)$ is a function of the time $a$ since the last reset event. We also call $a$ the age of the walker. The present work is an extension of previous studies where the resetting probability is time-independent. Using the renewal approach, we have established the formulations for the stationary occupation distribution and the MFPT between arbitrary two nodes, which are expressed in terms of the spectrum of the transition matrix and the modified transition matrix, and some resetting parameters. In particular, we consider two concrete time-dependent resetting protocols. The one is that $\phi(a)$ is step-shaped function of $a$, where the resetting probability switches from one value to another one as $a$ crosses a threshold value $a_c$. The other one is that $\phi(a)=\frac{a}{a+a_c}$ is a strictly increasing function of $a$. Both cases are exactly solvable. Finally, we demonstrate the theoretical results on a ring network and a Cayley tree for the two resetting protocols. We find that the MFPT can be further accelerated by the time-modulated resetting probability for a wide range of $a_c$, compared with the constant resetting probability. Therefore, time-modulated resetting protocols are more efficient in expediting the completion of a random search process than a simple constant-probability resetting protocol.

There are still some open questions concerning the resetting paradigm. In the future, it is worth studying other types of random walks under resetting, such as biased random walks \cite{PhysRevE.87.012112,PhysRevE.87.062140}, maximum entropy random walks \cite{PhysRevLett.102.160602}, and so on. Moreover, it would be also interesting to consider the factor of resetting costs on searching processes. In this context, how to find an optimal trade-off between the searching time cost and the resetting cost is a challenging issue \cite{arXiv:2112.11416}.

\appendix
\section{Spectral decomposition for transition matrix without resetting}\label{app1}
For standard random walks, the transition matrix can be written as $\bm{W}=\bm{D}^{-1} \bm{A}$, where $\bm{D}={\rm{diag}} \left\{ {{k_1}, \ldots ,{k_N}} \right\}$ is a diagonal matrix and $\bm{A}$ is the adjacency matrix of the underlying network. $\bm{W}$ can be rewritten as 
\begin{eqnarray}\label{a1.3}
\bm{W}=\bm{D}^{-1/2}{ \tilde{\bm{A}} }\bm{D}^{1/2}, 
\end{eqnarray}
where $\tilde{\bm{A}}=\bm{D}^{-1/2}\bm{ A}\bm{D}^{-1/2}$ is real-valued symmetric matrix for undirected networks ($\bm{A}=\bm{A}^\top$). Therefore, $\bm{W}$ is diagonalizable (i.e., spectral decomposition), and the eigenvalues of $\bm{W}$ and $\tilde{\bm{A}}$ are the same and are all real. Letting $\left| {{v _\ell}} \right\rangle$ denotes the right eigenvector corresponding to the $\ell$th eigenvalue of $\tilde{\bm{A}}$, it is not hard to verify that $| {{\phi _\ell}} \rangle  = {\bm{D}^{ - 1/2}}\left| {{v _\ell}} \right\rangle$ and $\langle {{{\bar \phi }_\ell}} | = \langle {{v _\ell}} |{\bm{D}^{  1/2}}$. 

The spectral decomposition for the transition matrix $\bm{W}$ is given by
\begin{eqnarray}\label{a1.1}
\bm{W} = \sum\limits_{\ell = 1}^N {{\lambda_\ell}| {{\phi _\ell}} \rangle } \langle {{{\bar \phi }_\ell}} |,
\end{eqnarray}
where $\lambda_\ell$ is the $\ell$th eigenvalue of $\bm{W}$, and the corresponding left eigenvector and right eigenvector are respectively $\langle {{{\bar \phi }_\ell}} |$ and $| {{\phi _\ell}} \rangle$, satisfying $\langle {{{\bar \phi }_\ell}} | {{\phi _m}} \rangle  = {\delta _{\ell m}}$ and $\sum_{\ell=1}^{N} |\phi_{\ell} \rangle \langle {\bar \phi}_{\ell}|=\bm{I}$.

Since $\bm{W}$ is a stochastic matrix, its maximal eigenvalue is equal to one. Without loss of generality, we let $\lambda_1=1$ and the values of other eigenvalues is less than one. Since the sum of each row of $\bm{W}$ is equal to one, the right eigenvetor corresponding to $\lambda_1=1$ is simply given by $| {{\phi _1}} \rangle  = {\left( {1,1, \ldots ,1} \right)^\top}$. The occupation probability $P^{0}_{ij}(t)$ without resetting is given by 
\begin{eqnarray}\label{a1.2}
{P^{0}_{ij}}(t) =  {\langle i } | {{\bm{W}^t}} | j \rangle = \sum\limits_{\ell = 1}^N {\lambda_\ell^t\langle i | {{\phi _\ell}} \rangle } \langle {{{\bar \phi }_\ell}} | j \rangle ,
\end{eqnarray}
where $| i \rangle $ denotes the canonical base with all its components equal to 0 except the $i$th one, which is equal to 1. In the limit of $t \to \infty$, all the eigenmodes decay to zero, except to the stationary eigenmode corresponding to $\lambda_1=1$. Therefore, we get to the occupation probability at stationary in the absence of resetting, ${P_j^{0}}(\infty ) = \langle {{{\bar \phi }_1}} | j \rangle$.

\section{Derivation of $\tilde Q_{ij}^0(s)$}\label{app2}
In the absence of resetting, the occupation probability and first passage probability satisfies the following relation, 
\begin{eqnarray}\label{a2.1}
{P^{0}_{ij}}\left( t \right) = {\delta _{t0}}{\delta _{ij}} + \sum\limits_{t' = 0}^t {{F_{ij}^{0}}\left( {t'} \right){P^{0}_{jj}}\left( {t - t'} \right)} ,
\end{eqnarray}
where $F_{ij}^{0}(t)$ is the first passage probability at time $t$ in the absence of resetting process.
In the Laplace domain, we have 
\begin{eqnarray}\label{a2.2}
{{\tilde F^{0}}_{ij}}\left( s \right) = \frac{{{{\tilde P^{0}}_{ij}}\left( s \right) - {\delta _{ij}}}}{{{{\tilde P^{0}}_{jj}}\left( s \right)}}.
\end{eqnarray}
In terms of Eq.(\ref{a1.2}), $\tilde{P}^{0}_{ij}(s)$ can be calculated as, 
\begin{eqnarray}\label{a2.3}
\tilde P_{ij}^0\left( s \right) = \sum\limits_{t = 0}^\infty  {{e^{ - st}}P_{ij}^0\left( t \right)}  = \frac{ \langle {{{\bar \phi }_1}} | j \rangle}{{1 - {e^{ - s}}}} + \sum\limits_{\ell = 2}^N {\frac{{\langle i | {{\phi _\ell}} \rangle \langle {{{\bar \phi }_\ell}} | j \rangle }}{{1 - {\lambda _\ell}{e^{ - s}}}}}  . \nonumber \\
\end{eqnarray}
Since $F_{ij}^{0}(t)=Q_{ij}^{0}(t-1)-Q_{ij}^{0}(t)$ for $t \ge 1$ and $F_{ij}^{0}(0)=1-Q_{ij}^{0}(0)$ for $t =0$, we have ${{\tilde F}_{ij}^{0}}\left( s \right) =1+ \left( {{e^{ - s}} - 1} \right){{\tilde Q}_{ij}^{0}}\left( s \right)$. Therefore, we have 
\begin{eqnarray}\label{a2.4}
\tilde Q_{ij}^0\left( s \right) = \frac{{1 - \tilde F_{ij}^0\left( s \right)}}{{1 - {e^{ - s}}}} = \frac{{\tilde P_{jj}^0\left( s \right) - \tilde P_{ij}^0\left( s \right) + {\delta _{ij}}}}{{\left( {1 - {e^{ - s}}} \right)\tilde P_{jj}^0\left( s \right)}}.
\end{eqnarray}
Substituting Eq.(\ref{a2.3}) into Eq.(\ref{a2.4}), we obtain
\begin{eqnarray}\label{a2.5}
\tilde Q_{ij}^0(s) = \frac{{\sum\limits_{\ell = 2}^N {\frac{{\langle j | {{\phi _\ell}} \rangle \langle {{{\bar \phi }_\ell}} | j \rangle  - \langle i | {{\phi _l}} \rangle \langle {{{\bar \phi }_l}} | j \rangle }}{{1 - {\lambda _\ell}{e^{ - s}}}}}  + {\delta _{ij}}}}{{\langle {{{\bar \phi }_1}} | j \rangle  + \left( {1 - {e^{ - s}}} \right)\sum\limits_{\ell = 2}^N {\frac{{\langle j | {{\phi _\ell}} \rangle \langle {{{\bar \phi }_\ell}} | j \rangle }}{{1 - {\lambda _\ell}{e^{ - s}}}}} }}.
\end{eqnarray}
Letting $s=0$ in Eq.(\ref{a2.5}), we obtain the mean first passage time in the absence of resetting,
\begin{eqnarray}\label{a2.6}
\langle {T_{ij}^0} \rangle & =& \tilde Q_{ij}^0\left( 0 \right) \nonumber \\ & =& \frac{1}{{\langle {{{\bar {\phi} }_1}} | j \rangle }}\left( {\sum\limits_{\ell = 2}^N {\frac{{\langle j | {{\phi _\ell}} \rangle \langle {{{\bar \phi }_\ell}} | j \rangle  - \langle i | {{\phi _\ell}} \rangle \langle {{{\bar \phi }_\ell}} | j \rangle }}{{1 - {\lambda _\ell}}}}  + {\delta _{ij}}} \right) .  \nonumber \\ 
\end{eqnarray}

\section{Derivation of $ Q_{ij}^0(t)$}\label{app3}
We will derive the expression of the survival probability in the absence of resetting process. There is a trap located at node $j$, and the walker starts from node $i$ at $t=0$. Let us denoted by $Q_{ij}^0(t)$ the survival probability, that is the probability that the walker survives up to time $t$. We first consider the case $i \neq j$. Let $\bm{W_j}$ be the matrix by letting all the entries in the $j$th row and the $j$th column of transition matrix $\bm{W}$ equal to zero. $Q_{ij}^0(t)$ can be expressed as
\begin{eqnarray}\label{a3.1}
Q_{ij}^0\left( t \right) = \sum\limits_{k = 1}^N {{{\left( \bm{W_j^t} \right)}_{ik}}}.
\end{eqnarray}
For the standard random walk, the matrix $\bm{W_j}$ can be written as $\bm{W_j} =\bm{D}^{-1} \bm{A_j}$, where $\bm{D}={\rm{diag}} \left\{ {{k_1}, \ldots ,{k_N}} \right\}$ is a diagonal matrix as before, and $\bm{A_j}$ is obtained by letting all the entries in the $j$th row and the $j$th column of the network adjacency matrix $\bm{A}$ equal to zero. It is not hard to prove that $\bm{W_j}$ is diagonalizable as $\bm{W_j}$ is conjugated to a real symmetric matrix $\bm{A_j}$. Letting $\zeta_\ell ^{(j)}$ be the $\ell$th eigenvalue of $\bm{A_j}$, and the associated eigenvector be $|{u_\ell^{(j)}}\rangle$, the spectral decomposition for $\bm{W_j}$ reads
\begin{eqnarray}\label{a3.2}
\bm{W_j} = \sum\limits_{\ell = 1}^{N } {\zeta _\ell^{(j)}} | {\psi _\ell^{(j)}} \rangle \langle {\bar \psi _\ell^{(j)}} |,
\end{eqnarray}
where $\langle {\bar \psi _\ell^{(j)}} |= \langle u_\ell^{(j)} | \bm{D}^{1/2}$ and $| {\psi _\ell^{(j)}} \rangle= \bm{D}^{-1/2} | u_\ell^{(j)} \rangle$ are respectively the left eigenvector and right eigenvector of $\bm{W_j}$ corresponding to the $\ell$th eigenvalue, satisfying $\langle {{{\bar \psi }_\ell^{(j)}}} | {{\psi _m^{(j)}}} \rangle  = {\delta _{\ell m}}$ and $\sum_{\ell=1}^{N} |\psi_{\ell}^{(j)} \rangle \langle {\bar \psi}_{\ell}^{(j)}|=\bm{I}$. 

According to Eq.(\ref{a3.2}), the survival probability can be computed by
\begin{eqnarray}\label{a3.3}
Q_{ij}^0(t) = \sum\limits_{k = 1}^{N } \sum\limits_{\ell = 1}^{N } ({\zeta _\ell^{(j)}})^t \langle i | {\psi _\ell^{(j)}} \rangle \langle {\bar \psi _\ell^{(j)}} | k \rangle, \quad  {\rm{for}} \quad i\neq j.  \nonumber \\ 
\end{eqnarray}

If the target is located at the starting node, $Q_{ii}^0(t)$ is the probability that the walker does not return to the original node. We define $Q_{ii}^0(t=0)=0$, and for $t \geq 1$,    
\begin{eqnarray}\label{a3.4}
Q_{ii}^0(t) = \sum\limits_{j = 1}^N {\sum\limits_{k = 1}^N {{W_{ij}}} } {\left( \bm {W_i^{t - 1}} \right)_{jk}}.
\end{eqnarray}
According to Eq.(\ref{a3.2}), $Q_{ii}^0(t)$ for $t\leq 1$ is written as 
\begin{eqnarray}\label{a3.5}
Q_{ii}^0(t) = \sum\limits_{j = 1}^{N } \sum\limits_{k = 1}^{N } \sum\limits_{\ell = 1}^{N } W_{ij} ({\zeta _\ell^{(i)}})^{t-1} \langle j | {\psi _\ell^{(i)}} \rangle \langle {\bar \psi _\ell^{(i)}} | k \rangle.
\end{eqnarray}

\begin{acknowledgments}
This work was supported by the National Natural Science Foundation of China (11875069, 61973001).
\end{acknowledgments}

\end{document}